\documentclass[sigconf,nonacm]{acmart}
\usepackage[T1]{fontenc}
\IfFileExists{libertine.sty}{}{\usepackage{lmodern}}

\usepackage{amsmath}
\usepackage{booktabs}
\usepackage{enumitem}
\usepackage{tikz}
\usetikzlibrary{arrows.meta, positioning, calc}

\begin{document}

\title{Understanding and Mitigating Training Instability in Large-Scale Recommender Systems: A Cascade Perspective}

\author{Yuening Li\texorpdfstring{\textsuperscript{*}}{*}, Zeyu Wang\texorpdfstring{\textsuperscript{*}}{*}, Zhen Zhang\texorpdfstring{\textsuperscript{*}}{*}, Jiayi Liu, Zhimeng Jiang, Yuji Roh, Weikai Xu, Mingyan Gao, Onkar Dalal, Chuan He, Liang Liu}
\affiliation{%
  \institution{Google}
  \country{}
}

\renewcommand{\shortauthors}{Y. Li, et al.}

\begin{abstract}
Training large-scale recommender systems on dynamic industrial platforms, such as Short-Form Video (SFV) services with billions of daily interactions, faces critical stability challenges. Rapid distribution shifts in user data and increasing model complexity frequently trigger a divergence cascade, typically initiating with \textbf{gradient explosion}, propagating to \textbf{weight explosion}, and culminating in \textbf{loss explosion}. This \textbf{G-W-L cascade} severely impedes model convergence and performance. This paper provides a systematic analysis of this G-W-L cascade within an industrial SFV recommender context. We then propose and evaluate a \textbf{structured mitigation framework} where interventions are explicitly mapped to counteract specific stages of the cascade: gradient (G) control, weight/activation (W) stabilization, and loss (L) shaping. Our contribution provides a structured framework for understanding and tackling training divergence, offering practical insights for deploying robust, continuously updated recommenders in large-scale, non-stationary settings. We empirically validate the effectiveness of these strategies on a major industrial SFV recommendation system.
\end{abstract}

\ccsdesc[500]{Information systems~Personalization}
\ccsdesc[500]{Information systems~Recommender systems}

\keywords{Recommender System; Optimization; Training Stability; Model Scaling}

\maketitle

\begingroup
\renewcommand\thefootnote{*}
\footnotetext{Equal contribution. Correspondence to: \texttt{yueningli@google.com}.}
\endgroup

\section{INTRODUCTION}
Modern Short-Form Video (SFV) platforms handle billions of user interactions daily, creating an extremely dynamic environment for training recommender systems. This scale and constant evolution bring major stability challenges. While standard fixes like gradient clipping \cite{tang2023improving, jiang2026addressing}, adaptive optimizers, learning rate schedules \cite{cohen2021gradient, gilmer2021loss}, and normalization layers are widely used, they often prove insufficient. Their effectiveness can be hit-or-miss, especially when faced with the \textbf{rapid data distribution shifts ($t$)} and pressures from \textbf{aggressive model scaling} common in continuous training on these platforms. These factors do not just make existing tools less reliable; they can actively trigger a domino effect of instability. For instance, simple gradient clipping might stop gradients from exploding but could also unintentionally flatten important learning signals, slowing down the model's ability to learn or adapt to new data. The sophisticated deep learning models used today, often with attention mechanisms \cite{vaswani2017attention}, combined with ever-changing input data, need more than generic fixes; they need an approach that understands the sequence of failure.

Much of the existing research offers solutions to training instability as isolated fixes. However, there is a missing piece: a deep dive into the common triggers and the specific chain reaction---from \textbf{G}radients to \textbf{W}eights to \textbf{L}oss (which we term the G-W-L cascade)---that leads to model divergence in these demanding, large-scale recommender systems. We often see instability starting with issues like data shifts or scaling efforts, which then set off this G-W-L cascade. Furthermore, the field lacks a clear framework that compares different solutions by mapping them to the specific stage of this cascade they address. This paper aims to fill these gaps. Our key contributions are:
\begin{itemize}[leftmargin=*]
    \item \textbf{Analyzing the Divergence Chain Reaction:} We conduct an empirical analysis of training dynamics in large-scale SFV recommender models. We identify common triggers (such as data distribution shifts and model scaling pressures) and detail the sequential links (Gradient $\rightarrow$ Weight $\rightarrow$ Loss explosion) within the instability cascade, which we formalize as the G-W-L cascade.
    \item \textbf{Structured Mitigation Framework:} Based on our analysis of the G-W-L cascade, we propose a clear way to categorize stability techniques. This framework maps solutions to the specific stage of the cascade they target: controlling gradients, stabilizing weights, or shaping the loss function.
    \item \textbf{Real-world Comparative Evaluation:} We present results from a large-scale empirical study on an industrial SFV recommender model. This study compares a variety of optimization and architectural changes, demonstrating a robust training approach that greatly improves both stability and the model's final performance by strategically addressing different parts of the G-W-L cascade.
\end{itemize}

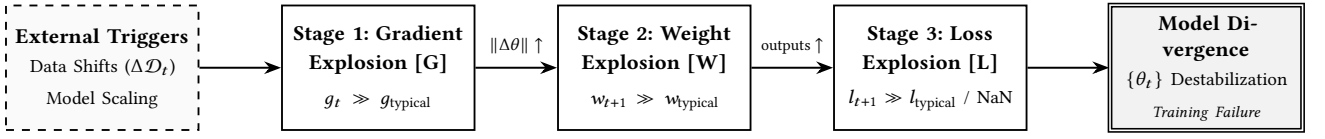
\begin{figure*}[t]
\centering
\begin{tikzpicture}[
    >=Stealth,
    box/.style={
        rectangle,
        draw=black,
        line width=0.7pt,
        fill=white,
        text width=2.3cm,
        minimum height=1.65cm,
        inner sep=3.5pt,
        align=center,
        font=\small
    },
    trig/.style={
        box,
        dashed,
        fill=black!2
    },
    fail/.style={
        box,
        double,
        line width=0.6pt,
        fill=black!5
    },
    arr/.style={
        ->,
        line width=0.75pt,
        draw=black
    },
    arrlbl/.style={
        font=\scriptsize,
        above=2pt,
        align=center
    }
]

    \node[trig] (trig) at (0, 0) {
        \textbf{External Triggers}\\[3pt]
        \footnotesize
        Data Shifts ($\Delta\mathcal{D}_t$)\\[2pt]
        Model Scaling
    };

    \node[box] (grad) at (3.65, 0) {
        \textbf{Stage 1: Gradient}\\[1pt]
        \textbf{Explosion [G]}\\[3pt]
        \footnotesize
        $g_t \gg g_{\mathrm{typical}}$
    };

    \node[box] (weight) at (7.30, 0) {
        \textbf{Stage 2: Weight}\\[1pt]
        \textbf{Explosion [W]}\\[3pt]
        \footnotesize
        $w_{t+1} \gg w_{\mathrm{typical}}$
    };

    \node[box] (loss) at (10.95, 0) {
        \textbf{Stage 3: Loss}\\[1pt]
        \textbf{Explosion [L]}\\[3pt]
        \footnotesize
        $l_{t+1} \gg l_{\mathrm{typical}}$ / NaN
    };

    \node[fail] (div) at (14.60, 0) {
        \textbf{Model Divergence}\\[3pt]
        \footnotesize
        $\{\theta_t\}$ Destabilization\\[2pt]
        \textit{\scriptsize Training Failure}
    };

    \draw[arr] (trig) -- (grad);
    \draw[arr] (grad) -- node[arrlbl] {$\|\Delta \theta\| \uparrow$} (weight);
    \draw[arr] (weight) -- node[arrlbl] {outputs $\uparrow$} (loss);
    \draw[arr] (loss) -- (div);

\end{tikzpicture}
\caption{The G-W-L Cascade: A sequential failure chain reaction leading to model divergence. External perturbations (data distribution shifts $\Delta\mathcal{D}_t$ or aggressive scaling) trigger gradient explosion ($g_t \uparrow$), inducing large parameter updates ($\|\Delta\theta\| \uparrow$) that cause weight explosion ($w_{t+1} \uparrow$). Inflated model weights produce extreme network outputs, driving loss explosion ($l_{t+1} \uparrow$) and culminating in complete model divergence.}
\label{fig:cascade}
\end{figure*}

\section{METHODOLOGY}

\subsection{Model Divergence Problem Definition}
When training large recommender systems, model divergence is a serious roadblock. It happens when the learning process goes off track, causing the model's settings (parameters) to not settle down. This leads to unreliable predictions, bad recommendations, and a worse experience for users. Factors like trying to scale the model too quickly or changes in the input data over time ($t$) can make this problem more likely or more severe.

Let $\mathcal{M}$ be a large sequential recommender model with parameters $\theta$. We train it using a dataset $\mathcal{D} = \{s_i\}_{i=1}^N$ of user activity sequences, a loss function $\mathcal{L}$ to measure errors, and an iterative learning algorithm. At each training step $t$, we watch three key numbers: the size of the gradient (how much to change the model) $g_t = \|\nabla \mathcal{L}(\theta_t)\|_2$, the size of the model's parameters $w_t = \|\theta_t\|_2$, and the training error $l_t = \mathcal{L}(\theta_t)$. Divergence usually shows up when one or more of these numbers shoot up unexpectedly:
\begin{itemize}[leftmargin=*, nosep]
    \item \textbf{Gradient Explosion}: $g_t$ quickly grows out of control.
    \item \textbf{Weight Explosion}: $w_t$ also rapidly grows out of control.
    \item \textbf{Loss Explosion}: $l_t$ suddenly jumps to a very high value.
\end{itemize}
Our goal is to understand how $g_t, w_t,$ and $l_t$ affect each other, especially how they lead to an unstable training process, and to find reliable ways to stop this from happening.

\subsection{The G-W-L Divergence Cascade}
We observe a common pattern of events, which we call the G-W-L cascade, that leads to model divergence. This is shown in Figure \ref{fig:cascade}. The trouble often starts when external factors, such as significant \textbf{Data Distribution Shifts} at time $t$ or aggressive \textbf{Model Scaling}, put stress on the training process. This can trigger a \textbf{Gradient Explosion} ($g_t \uparrow$), where the gradient's size at step $t$ becomes far too large:
\begin{equation}
g_t = \|\nabla \mathcal{L}(\theta_t)\|_2 \gg g_{\mathrm{typical}}
\end{equation}
The standard way to update model parameters is $\theta_{t+1} = \theta_t - \eta \nabla \mathcal{L}(\theta_t)$ (where $\eta$ is the learning rate). This formula shows that very large gradients $\nabla \mathcal{L}(\theta_t)$ can cause overly large changes to the weights, meaning $\|\theta_{t+1} - \theta_t\|$ becomes big. These large changes can then cause a \textbf{Weight Explosion} ($w_{t+1} \uparrow$), where the overall size of the model's parameters grows uncontrollably:
\begin{equation}
w_{t+1} = \|\theta_{t+1}\|_2 \gg w_{\mathrm{typical}}
\end{equation}
As a result, these inflated weights $\theta_{t+1}$ often make the network produce extreme outputs for the data it sees. This, in turn, causes the loss value $l_{t+1}$ to shoot up, a problem we call \textbf{Large Network Outputs \& Loss Explosion} ($l_{t+1} \uparrow$). The loss might even become infinity or NaN (Not a Number):
\begin{equation}
l_{t+1} = \mathcal{L}(\theta_{t+1}) \gg l_{\mathrm{typical}}
\end{equation}
This whole chain reaction ends in \textbf{Model Divergence}. At this stage, the learning process completely breaks down: the model's parameters $\{\theta_t\}$ do not settle into a stable or useful state. In practice, we usually detect divergence when $g_t$, $w_t$, or $l_t$ explode. The sections below discuss ways to tackle each stage of this G-W-L cascade.

\subsection{Mitigating Gradient Explosion (Controlling \texorpdfstring{$g_t$}{g\_t})}
To stop gradient explosion, we can use methods that change how gradients are calculated or used for updates. These techniques adjust the gradient information. Instead of the standard update, we use a more general form:
\begin{equation}
\theta_{t+1} = \theta_t - \eta f(\nabla \mathcal{L}(\theta_t))
\label{eq:grad}
\end{equation}
Here, $f(\cdot)$ is a function that transforms the raw gradient. Examples include Adaptive Gradient Clipping (like Clippy \cite{tang2023improving}), smoother activation functions in the model (like SiLU \cite{elfwing2018sigmoid} or SmeLU \cite{shamir2020smooth}), layers that normalize values (like RMSNorm \cite{zhang2019root}), and optimization algorithms that have built-in normalization (like Lion \cite{chen2024symbolic}).

\subsection{Mitigating Weight Explosion (Controlling \texorpdfstring{$w_{t+1}$}{w\_\{t+1\}})}
Weight explosion is usually handled by methods that directly limit the size of the parameters or how much they can change. These methods apply a constraint or projection $P(\cdot)$ to the parameters after they have been updated:
\begin{equation}
\theta_{t+1} = P(\theta_t - \eta f(\nabla \mathcal{L}(\theta_t)))
\end{equation}
In this equation, $f(\nabla \mathcal{L}(\theta_t))$ represents the gradient's influence, possibly already adjusted as in Eq. (\ref{eq:grad}). Common techniques are Value Clipping and Activation Clipping/Normalization \cite{krizhevsky2010convolutional}.

\subsection{Mitigating Loss Explosion (Controlling \texorpdfstring{$l_{t+1}$}{l\_\{t+1\}})}
Loss explosion can be managed by using loss functions that are not as heavily affected by large prediction errors. This means swapping out the usual loss function $\mathcal{L}$ for a modified, more stable one, $\mathcal{L}'$:
\begin{equation}
l_{t+1} = \mathcal{L}'(\theta_{t+1})
\end{equation}
A common strategy here is to use Robust Loss Functions, such as Huber Loss \cite{huber1992robust}.

\section{EXPERIMENTS}
This section details experiments designed to investigate the G-W-L explosion cascade and evaluate the efficacy of targeted interventions. Our inquiry centers on two Research Questions (RQs):

\textbf{RQ1: Cascade Interdependence:} To what extent do gradient, weight, and loss metrics exhibit co-occurring explosions during divergence episodes, and how do interventions targeting one component of the hypothesized G-W-L cascade affect the others?

\textbf{RQ2: Training Factor Influence:} How do key training factors (optimizer, architectural choices, loss functions), specifically those designed to modulate gradients (G), internal model values/weights (W), or loss characteristics (L), impact the stability (occurrence and magnitude of explosions) and end-task performance of large-scale recommender models?

\subsection{Experimental Setup}
We evaluated the following configurations on a large-scale industrial recommender model:
\begin{itemize}
    \item \textbf{Baseline}: Control setup using standard practices (e.g., Adam optimizer \cite{kingma2014adam}, basic learning rate schedule) without specialized stability interventions.
    \item \texttt{\textbf{G}} \textbf{(Gradient Control)}: Enhances gradient stability via SmeLU activation, RMSNorm in attention layers, and global gradient clipping.
    \item \texttt{\textbf{W}} \textbf{(Weight/Activation Control)}: Stabilizes internal network states using activation value clipping in dense linear layers.
    \item \texttt{\textbf{L}} \textbf{(Loss Control)}: Improves loss stability by replacing Mean Squared Error (MSE) with Huber loss for designated regression tasks.
    \item \texttt{\textbf{G+W+L}} \textbf{(Combined Control)}: Integrates all strategies from \texttt{G}, \texttt{W}, and \texttt{L} for a comprehensive stability approach.
\end{itemize}
We assessed performance using standard end-task metrics. For the binary classification tasks (BC T1 and BC T2), we used the Area Under the Curve (AUC), which quantifies the overall ability of the model to distinguish between two classes. Mathematically, AUC corresponds to the area under the Receiver Operating Characteristic (ROC) curve, which plots the true positive rate ($\text{TPR}$) against the false positive rate ($\text{FPR}$) at various classification thresholds, and can be approximated as $\text{AUC} = \int_0^1 \text{TPR}(\text{FPR}^{-1}(x))\,dx$. For the regression task (Reg T1), we used the Root Mean Squared Error (RMSE), which measures the average magnitude of the prediction errors and is defined as $\text{RMSE} = \sqrt{\frac{1}{n} \sum_{i=1}^n (\hat{y}_i - y_i)^2}$, where $\hat{y}_i$ is the predicted value, $y_i$ is the true value, and $n$ is the number of data points. Lower RMSE indicates better predictive performance. In addition to these metrics, we tracked internal model dynamics, specifically the gradient norms $g_t$ and the loss values $l_t$ over time.

\subsection{Results and Discussion}

\subsubsection{RQ1: Cascade Interdependence}
While our aggregated snapshot data (Tables \ref{tab:end_task} and \ref{tab:internal_dynamics}) does not permit a fine-grained temporal validation of the $g_t \uparrow \implies w_{t+1} \uparrow \implies l_{t+1} \uparrow$ sequence for every divergence event (RQ1), the results provide compelling evidence for the interconnectedness of these components. Specifically, interventions targeting one element of the G-W-L cascade always influenced the others. For instance, as detailed under RQ2, strategies aimed at the loss component (L) significantly affected gradients (G), while gradient interventions (G) had repercussions for weights (W) and loss (L). This supports the hypothesis of a tightly coupled G-W-L cascade over isolated failure points.

\begin{table}[htbp]
\centering
\caption{End-Task Performance: Relative Percentage Change from Baseline. AUC: positive \%\ensuremath{\Delta} is better. RMSE: negative \%\ensuremath{\Delta} is better.}
\label{tab:end_task}
\footnotesize
\begin{tabular}{lrrr}
\toprule
\textbf{Design Name} & \textbf{BC T1} & \textbf{BC T2} & \textbf{Reg T1} \\
 & \textbf{(AUC \%\ensuremath{\Delta})} & \textbf{(AUC \%\ensuremath{\Delta})} & \textbf{(RMSE \%\ensuremath{\Delta})} \\
\midrule
Baseline & 0.00 \% & 0.00 \% & 0.00 \% \\
Treatment Gradient & 0.05 \% & 0.19 \% & -0.13 \% \\
Treatment Weight & 0.03 \% & 0.06 \% & -0.09 \% \\
Treatment Loss & 0.21 \% & 0.38 \% & 0.16 \% \\
Treatment All & 0.25 \% & 0.38 \% & -0.03 \% \\
\bottomrule
\end{tabular}
\end{table}

\subsubsection{RQ2: Training Factor Influence}
\textbf{A. Impact on End-Task Performance (Relative Change from Baseline)}

As shown in Table \ref{tab:end_task}, interventions targeting G, W, and L components influenced end-task performance relative to the Baseline. The comprehensive \texttt{G+W+L} strategy consistently enhanced performance across all tasks, delivering the highest AUC gains in BC T1 (0.25\%) and matching the best in BC T2 (0.38\%), alongside a slight RMSE reduction in Reg T1 (-0.03\%). The \texttt{L} intervention yielded substantial AUC improvements (0.21\% for BC T1, 0.38\% for BC T2), though it slightly increased RMSE in Reg T1 (0.16\%).

\textbf{B. Impact on Internal Training Dynamics ($g_t, w_t, l_t$) (Relative Change from Baseline)}

Table \ref{tab:internal_dynamics} presents the relative changes in gradient norms ($g_t$) and total loss ($l_t$) to directly assess each intervention's influence on the G-W-L components. Notably, the Baseline configuration exhibited high initial $g_t$ and $l_t$ values (based on observations not detailed in the relative changes table), indicative of an inherently unstable training regime.

\begin{table}[htbp]
\centering
\caption{Internal Training Dynamics: Relative Percentage Change from Baseline. $g_t$ is the L2 norm of non-embedding parameter gradients (before clipping); $l_t$ is the total training loss.}
\label{tab:internal_dynamics}
\footnotesize
\begin{tabular}{lrr}
\toprule
\textbf{Design Name} & \textbf{$g_t$ (\% \ensuremath{\Delta})} & \textbf{$l_t$ (\% \ensuremath{\Delta})} \\
\midrule
Baseline & 0.0 \% & 0.0 \% \\
Treatment Gradient & -63.7 \% & -3.6 \% \\
Treatment Weight & -19.6 \% & -7.2 \% \\
Treatment Loss & -66.7 \% & -62.5 \% \\
Treatment All & -83.6 \% & -64.8 \% \\
\bottomrule
\end{tabular}
\end{table}

The ablation study in Table \ref{tab:internal_dynamics} reveals how these targeted interventions affected key internal metrics, underscoring the G-W-L cascade hypothesis and the interdependence discussed in RQ1. The significant relative changes from the unstable Baseline further highlight the efficacy of these strategies:
\begin{itemize}
    \item \texttt{\textbf{G}}: The \texttt{G} intervention effectively controlled gradient magnitudes ($g_t$ -63.7\%). This complexity may stem from the multifaceted nature of the gradient-focused measures (SmeLU, RMSNorm, global clipping) or optimizer dynamics.
    \item \texttt{\textbf{W}}: Controlling activation magnitudes (\texttt{W}) successfully reduced not only weight norms ($w_t$ -11.6\%) and loss ($l_t$ -7.2\%), but also raw gradients ($g_t$ -19.6\%). This demonstrates effective W$\rightarrow$L and W$\rightarrow$G feedback, where more stable internal states promote overall training stability.
    \item \texttt{\textbf{L}}: Employing Huber loss (\texttt{L}) dramatically reduced both total loss ($l_t$ -62.5\%) and raw gradients ($g_t$ -66.7\%). This provides compelling evidence for a potent L$\rightarrow$G feedback loop: a robust loss function curbs instability at its source, leading to smaller, more stable gradients.
    \item \texttt{\textbf{G+W+L}}: The comprehensive \texttt{G+W+L} strategy yielded the most significant reductions in both raw gradients ($g_t$ -83.6\%) and total loss ($l_t$ -64.8\%). This equilibrium correlates with the observed improvements in end-task performance.
\end{itemize}

\subsection{Robust Divergence Prevention via a Holistic G+W+L Strategy}
Preventing divergence in large-scale models is vital, though direct pre-failure risk assessment is challenging. We thus used proxy indicators---internal clipping activity and extreme condition stress tests (e.g., high learning rates, unstable architectures)---to assess stability. Our comprehensive G+W+L strategy (integrating techniques including optimized learning, stable architectures, and robust activations) markedly improved internal stability. Rigorously validated via 12 A/A arms, the strategy confirmed neutral offline/online results and delivered a >28,000-fold reduction in key clipping indicators under standard conditions (when compared to diverged trainers), without harming end-task performance.

The G+W+L strategy's resilience was further proven under severe stress. In scenarios designed to induce failure (e.g., five-fold learning rate increases or inherently unstable architectural patterns) where baseline models consistently diverged, our approach ensured successful training completion. Moreover, it maintained significantly lower internal clipping activity, with improvements ranging from over 20-fold to more than 70-fold relative to baselines near their divergence point. This demonstrates that a proactive, multi-faceted G+W+L approach, guided by empirical stability indicators, confers substantial resilience against training divergence. Such resilience is critical for deploying and maintaining robust, large-scale models in dynamic industrial environments, effectively mitigating a common and costly failure mode.

\section{CONCLUSION}
This paper investigated training instability in large-scale industrial recommender systems through the lens of a Gradient $\rightarrow$ Weight $\rightarrow$ Loss (G-W-L) explosion cascade. Our contributions are threefold: characterizing this cascade, proposing a structured mitigation framework targeting its components, and providing an extensive comparative evaluation of these strategies on a real-world Short-Form Video recommender model. Our experiments demonstrate that interventions specifically designed to control gradient magnitudes (G), internal model values/weights (W), and loss characteristics (L) significantly influence training dynamics and model performance.

\bibliographystyle{ACM-Reference-Format}
\bibliography{references}

\end{document}